\documentclass{iaus260}
\usepackage{graphicx}
\pubyear{2009}
\volume{260}  
\pagerange{10--17}
\jname{The R\^ole of Astronomy in Society and Culture}
\editors{D. Valls-Gabaud \& A. Boksenberg, eds.}

\title["Are we alone ?" in different cultures]                   
{The question "Are we alone ?" in different cultures}  

\author[Jean Schneider]           
{Jean Schneider$^1$}
\affiliation{$^1$LUTh - Paris Observatory, 92190 Meudon, France, Jean.Schneider@obspm.fr}

\begin{document}
\maketitle

\begin{abstract}
A survey of the worldwide litterature reveals that the question ''Are we alone
in the Universe ?'' has been formulated only in the ''western'' litterature.
Here I try to understand why it is so. To investigate this problem it
is first necessary to clarify what ''western'' culture means.
\end{abstract}

\firstsection 
\section{Introduction}
The questions ''Is there life in the Universe outside Earth ?'' or
''Are we alone in the Universe ?'' has become one the main
drivers of Space Agencies around the world. See for instance 
the Cosmic Vision programme of the European Space Agency and
the NASA programme ''Origins''.
They are often claimed to be ''as old as Humanity itself''.
They indeed look very natural since Life is spread out over the whole Earth 
and therefore even a child
rising his eyes toward the sky can ask ''is there life out there ?''.
But, very surprisingly, there is no written occurence of this question
in ''non western'' ancient cultures.
In a first part of this paper I justify this statement. Then I will try
to understand why it so. I will thus be led to first clarify
what can characterize and delimitate ''western'' culture.
Then I will propose a hypothesis to explain why the question
of Life in the Universe has not been raised by non-western cultures.
Finally I will address the question ''why did this movement start in Greece ?''.

There are generally two ways to consider the question
of extraterrestrial life:  the point of view of living organisms, 
leading to the question ''Is there Life elsewhere in the Universe ?'', 
which is the subject of
 exobiology and extraterrestrial intelligence, leading to
the question ''Are we alone ?'' or ''Is there anybody out there ? '', 
which is subject of SETI (Search for Extra-Terrestrial Intelligence).
Also a connected question is the nature of Life: how different can it be from terrestrial life?
This question is symbolized by the word "Aliens" often found in the literature.
Here I will treat these three questions as if they were only one.

\section{Survey of the world-wide litterature and traditions}

The question of extraterrestrial life in the litterature since
the Greeks has been compiled in the remarkable books ''The Extreterrestrial
Life Debate 1750-1900 - The Idea of a Plurality of Worlds from Kant to Lowell'' \cite[Crowe (1986)]{Crowe1986}
and ''The Extraterrestrial life debate, antiquity to 1915''
  \cite[Crowe (1986)]{Crowe2008}. They are a must on this topic.
They represent an almost exhaustive compilation of all authors
having expressed an opinion on this debate. According to the name
index, about 600 authors are cited and, with only one exception, I have
never found a French author before 1900, who was not referenced in these books.

It is remarkable that almost all authors entering the debate have expressed 
that the existence of extraterrestrial life seemed natural to them.
Among the most famous authors, the only few remarkable exceptions are 
Aristotles, Augustine, Hegel, Schopenauer and to some
extent Plato. That means a very few skeptics among hundreds of optimists.
It is also strange that, while the debate has increased in intensity among the scientific
community at the end the the XIX$^{th}$ century, almost no philosopher after
Schopenauer was interested in this subject. Only H. Bergson in his {\it Evolution
Cr\'eatrice} and more vaguely C.S. Peirce and W. James did mention
the question of extraterrestrial life. 
It cannot be explained by ignorance: many novelists like Charles Cros, H.G. Wells, A. Strindberg\footnote{In his drama ''Father'', one the key characters worked on panspermia.},
Marconi, Stepledon \footnote{Olaf Stapledon (1886-1950), a british psychologist, envisaged communication with extraterrestrial in his "Last and First Men" (\cite[Stapledon (1930)]{Stapledon}).} and Tristan Bernard \footnote{French humorist, 1866-1947} did contribute to an outreach of the extraterrestrial life debate in the general culture.
Only in the second half of the XX$^{th}$ century
Paul Watzlawick, from the Palo Alto school in sociology, addressed seriously
the question of
communication with extraterrestrials (\cite[Watzlawick (1976)]{Watzlawick}).

The most important, although obvious, observation from Crowe's books is that all authors
cited are Europeans and (after 1800) North-Americans. It does not
seem to result from the limitation of the author's enquiry. No reference
to extraterrestrial life exists in "Astronomy Across Cultures - The History on Non-Western Cultures" 
\cite[Selin (2000)]{Selin} nor in "L'Astronomie des Anciens" \cite[Naze (2009)]{Naze}.
There seems to be an apparent exception in Jewish litterature: Moise Maimonides (circa 1135 - circa 1204) in the "Guide for the Perplexed" says
"The whole mankind at present in existence $[$...$]$ and every other species of animals, 
form an infinitesimal portion of the
permanent universe $[...]$  it is of great advantage that man should know his station, and not erroneously imagine that the
whole universe exists only for him." (Chapter XII p. 268).
But Maimonid was a European Jew living in Cordoba (Spain). He knew well ancient Greeks' 
work and participated in the cultural atmosphere also represented by
Michael Scot (1175 - 1235) and Albertus Magnus (1193 - 1280) for instance 
who were among the Middle Age philosophers
supporting the idea of extraterrestrial life.

To be complete, one must say that there are references to non-human being in some of these
cultures, but they are rather of the "surnatural" angelic type. 

What is even more curious is that this question "why?" has never been discussed,
at least to my knowledge.

\section{Why does the extraterrestrial life debate exist only in ''western'' culture ?}
Here I will illustrate my argumentation by historical examples.
My purpose is nevertheless not a historical perspective. It is rather a-historical
and structural. I will develop  a hypothesis which rests on a main guiding principle:
"elsewhere" and "aliens" require some distanciation, some differentiation.
This principle is an {\it a priori} reading grid, somehow schematic compared to the complexity
of historical situations, like the galilean inertia principle apparently contradicted
by everyday life dominated by dissipative frictions.

There are two types of distanciation: the distianciation of concepts from
their empirical objects and the spatial distanciation. 
These two aspects are closely connected and in particular spatial distanciation
requires distianciation by concept as a prerequisite.
Let us nevertheless shortly discuss them separately.\\

{\it Conceptual distanciation}

The idea that life can exist \underline{elsewhere} requires that the word "Life"
is not consubstancial with the living beings with which we have personal relationships.
In other words, it requires a \underline{concept} of "Life". Only concepts can be generalized.
This points toward the "universalizing" structure of concepts. What is called "abstraction" is
then the result of this universalization.\\

We can at this point try to charaterize "Western culture" as the culture of concepts 
with their mathematization and the constraints that they impose.

Concepts are created by the words naming them: see the ideas of nominalism (Abelard) and 
the Berkeleysian so-called idealism \footnote{The truth is that the so-called materialsim is in
fact a true idealism as we never experience anything like "matter itself", but only perceptions and what language makes of it.}.
Moreover, what is not subject of language cannot be imagined different: to imagine that things are different 
one must give them names AND detach the word from the designated object. Hence the above-mentionned 
conceptual distanciantion.
An example is given by the idea of "circle": it is an abstraction insofar as
there is no perfect circle in nature \footnote{See {\it The Origin of Geometry} by E. Husserl.}
and a source of universalization since it allows to put all empirical curves ressembling a circle
into the same single category.
Another, less abstract, example of universalization is given by the introduction
of the metrical system which abandonned local customs for a "universal" length unit
(the corresponding "universe" being the Earth, shared by every country).

The latter example is a good transition toward spatial distanciation.

{\it Spatial distanciation}

Euclid's Elements  introduced a rigorous structuration
of spatiality, the realm of potential freedom of motion.
An important consequence was Thales' theorem. The latter  permitted one to
make rigorous statements on objects (their length) inaccessible to 
direct manipulation. As such, it opened the possibility of
{\it extra}-polation, the possibility of transfering to distant
objects characteristics of objects within our reach, like harboring life
for "other worlds". It is also worthwhile to note that
the idea of proportion underlying Thales' theorem is
in Latin the same word as "reasoning" ("ratio"), another aspect of the above-mentionned
conceptual distanciation.
Moreover Euclid's geometry introduced homogeneity of space,
opening the possibility that 
"here" is not a center, not "the" center. It is not necessary to recall
 the fortune of this idea
with the end of geocentrism introduced by Aristarchus of Samos and Copernicus.
About the latter, it is interesting to note that there no reference to
extraterrestrial life in his writings.
In other words, one is thus led from distanciantion to 
 decentration.\\

This homogeneity underlines the great difference with Aristoteles'
conception of space for whom the Universe was divided into the 
Earth (the sublunar world) and Heavens (the superlunar world).
Both were very heterogeneous and it would have been illogical
to transfer to the Heavens something like terrestrial living organisms.

This rationalised structuration of space is significantly opposed
to the idea of  Ying and Yang where every "yang-like" notion contains
some some "ying-like" quality and vice versa. This Ying-Yang structure
is impossible to express in geometrical terms \footnote{It can nevertheless be mathematized in modern terms
thanks for instance to "non-well founded" set theory (J. Barwise and J. Etchemendy. "The Liar" Oxford University Press 1987)   or to Combinatory Logic (Schneider J. 
"La non-tratification" in {\it La psychanalyse et la r\'eforme de l'entendement} available at
http://www.obspm.fr/$\sim$schneider ).}. This difference in the treatment of space in ancient China
and Europe is well illustrated by the difference between chinese painting and italian perspective.


To summarize, our hypothesis is that the apparition of the 
theme of extraterrestrial life in the
Indo-European area, and its culmination in Greece, is related to 
the apparition of Euclidean geometry and of the so-called Greek logos.
\section{A societal note: conceptual distanciation and democracy}
In addition to distanciation, another important aspect of concepts is that
they are likely to be shared by every individual. Indeed, it belongs to the essence of concepts that
they are not the property of a political power. It results that the political power (King, Emperor) 
cannot be the source of concepts. They are are their own, impersonal, source. To express it in a radical way, they
\underline{are} their own power. In Astronomy, things were very different in
ancient China where, for instance, the few astronomical knowledge like the prediction of eclipses
or even the calendar were the private property of the Emperor, because they did provide some power.
In addition, concepts are open to debate. That is why concepts and democracy go together,
if by democracy one means "public debate" rather than such or such election systems.
And it is a fact that in this sense democracy has appeared in the part of the world in which  
the {\it logos} also appeared.

It is also interesting to note that in the European Age of Enlightenment where the extraterrestrial 
life debate gained in intensity with authors like Fontenelle, the idea of decentration gained also a societal
tone. This is for instance witnessed by Montesqieu's and Voltaire's work.\footnote{Montesiquieu: "If I knew a thing useful to me but harmful to my family, I would reject it. If I knew a thing useful to my family, but 
useless to my homeland, I would forget about it. If I knew a thing useful to my homeland or to Europe, but prejudicial to the human gender, I would consider it as a crime." in his {\it Carnets}. See also Voltaire's 
"point of view from Sirius" in his {\it Micromegas}.}. There is here a significant contrast with one of the old China's name: "The Empire of the Middle".

One may wonder if such considerations do not lead to a Europeocentrism.
Such a potential Europecentrism seems to culminate with Kant when he writes in his
{\it Idea for a Universal History from a Cosmopolitan Point of View}:
".. our continent [Europe] (which will probably give law, eventually, to all the others)..." 
(9th Thesis) \footnote{He meant
ethical laws, pointing toward human rights.}
 But to this real concern one can reply:

- that it is not an ideological position but a matter of fact that the entire world has
adopted the scientifico-technical concepts.

- that these concepts are not the only respectable values. For instance hospitality 
seems to be more developed today in non-European parts of the world. And notions like 
Ying and Yang are more useful in some human affairs than rigid rationality. The german
philosopher Heidegger has lengthly developed in his article {\it Dialogue with a Japanese}
 (in {\it On the Way to Language })
that western philosophy has a great deal to learn from the Japanese notion
of {\it koto ba} (which means something like "gracefulness").

In another vein, Greeks' literalism missed the kabbalistic approach of the reading of great texts
which  is undoubtedly one of the sources of psychoanalysis.

Note finally that if Chinese did have a somewhat elaborated technique, Greeks did
not have a systematic development of  technology. For instance, they 
used steam machines to open the heavy doors of they temples, but
did not think of applying it in a systematic way to everyday practical life
and therefore missed premises of industrialization.
\section{Why did all this start essentially in Greece ?} 
This movement did start in the Indo-European arc (which comprises arabic countries).
But it exploded in Greece a few centuries B.-C. . One could search for some geographical,
economical or climatic reason for that. But my thesis is that this greek
geographical location
is causeless. Its origin is pure genuine fortuitness, spontaneous generation.
This claim results from a "psychological theorem" according to which
ideas emerge from nowhere. This "theorem" is illustrated by the {\it a priori}
essence of concepts pointed out by Kant: concepts do not emerge FROM experience,
they are a prerequisite to make it intelligible. In another domain, 
modern language theories rest
on de Saussure's principle of {\it arbitrariness} of signs: linguistics
symbols are also given {\it a priori}.

\section{Conclusion}
The thesis presented here is open to debate. Disagreement with
the present views is of course 
always possible,
but any disagreeing opinion should at least  offer an alternative explanation of
the fact pointed out here that the extraterrestrial life debate seems to be restricted
to "western" litterature.

\end{document}